\documentclass[prc,aps,twocolumn,groupedaddress,floatfix,preprintnumbers]{revtex4}
\usepackage{amsmath,amssymb,multirow,epsfig,bm,color,hhline}
\usepackage[utf8]{}

\newcommand{\etal}{{\it et al.,\;}}

\newcommand{\beq}{\begin{equation}}
\newcommand{\eeq}{\end{equation}}
\newcommand{\bea}{\begin{eqnarray}}
\newcommand{\eea}{\end{eqnarray}}

\newcommand{\ssec}[1]{\section{#1}}
\newcommand{\Ethresh}{E_\text{thresh}}
\newcommand{\Ecm}{E_\text{c.m.}}

\newcommand{\itemovie}[2]{\item{File: \texttt{#1}\newline YouTube: {\href{#2}{\texttt{#2}}}}}
\newcommand*\sepline{%
  \begin{center}
    \rule[1ex]{.4\textwidth}{.6pt}
  \end{center}
}

\begin{document}

\title{
Pairing dynamics and solitonic excitations in collisions of medium-mass, identical nuclei
}

\author{Piotr Magierski$^{1,2}$}\email{piotr.magierski@pw.edu.pl}
\author{Andrzej Makowski$^{1}$} 
\author{Matthew C. Barton$^{1}$}
\author{Kazuyuki Sekizawa$^{3,4}$}
\author{Gabriel Wlaz{\l}owski$^{1,2}$}\email{gabriel.wlazlowski@pw.edu.pl}

\affiliation{$^1$Faculty of Physics, Warsaw University of Technology, Ulica Koszykowa 75, 00-662 Warsaw, Poland}
\affiliation{$^2$Department of Physics, University of Washington, Seattle, Washington 98195--1560, USA}
\affiliation{$^3$Department of Physics, School of Science, Tokyo Institute of Technology, Tokyo 152-8551, Japan}
\affiliation{$^4$Nuclear Physics Division, Center for Computational Sciences, University of Tsukuba, Ibaraki 305-8577, Japan}

\begin{abstract} 
We present results of collisions of $^{90}$Zr+$^{90}$Zr and $^{96}$Zr+$^{96}$Zr obtained within time-dependent density functional theory (TDDFT) extended to 
superfluid systems, known as time-dependent superfluid local density approximation (TDSLDA).
We discuss qualitatively new features occurring in collisions of two superfluid nuclei at energies in the vicinity of the Coulomb barrier. 
We show that a \textit{solitonic excitation}---an abrupt pairing phase distortion---reported previously [P.~Magierski et al., Phys. Rev. Lett. \textbf{119}, 042501 (2017)], increases the barrier for capture generating effective repulsion between colliding nuclei. 
Moreover we demonstrate that pairing field leads to  qualitatively different dynamics at the Coulomb barrier which manifests itself in a slower evolution of deformation towards a compact shape.
Last but not least, we show that magnitude of pairing correlations can be dynamically enhanced after collision. 
We interpret it as a dynamically induced $U(1)$ symmetry breaking, which leads to large-amplitude oscillations of pairing field and bear similarity to the pairing Higgs mechanism.
\end{abstract}

\maketitle

\ssec{Introduction} 
Pairing correlations in nuclear systems are one of the best known characteristics of non magic atomic nuclei~\cite{ring, dean, BrinkBroglia}. 
Various features related to high spin phenomena \cite{sorensen73,voigt83,shimizu89} or to large amplitude collective motion~\cite{barranco90, nakatsukasa16}, e.g. fission~\cite{bulgac16, bender20, bulgac20}, indicate that these correlations are crucial for our understanding of nuclear structure and dynamics.
There is, however, a subtle difference between pairing correlations and a superfluid phase.
In finite systems the distinction between the two is not easy to make. 
Pairing in atomic nuclei is usually theoretically described on a mean-field level, where the concept of pairing field plays the key role. 
It implicitly assumes the existence of superfluid phase described by the complex field playing the role of the order parameter.
Although the average magnitude of this field is an important ingredient of any theoretical description of medium or heavy nuclei, the other features related to this degree of freedom are usually omitted in the context of nuclear dynamics.
These features include spatial modulations (oscillations) of the order parameter, where both the magnitude and the phase may vary in space and time. 
It is well known that degrees of freedom of the pairing field are responsible for phenomena that abound in various superfluid condensed matter systems~\cite{barone,buzdin05,casalbuoni04}. 
Still, in atomic nuclei they are believed to play a minor role and the pairing field is usually reduced in nuclear physics to a single number defining the energy gap at the Fermi surface.
There are two arguments that justify such restricted treatment of nuclear superfluidity. 
The first one is related to the size of a nuclear Cooper pair, which in atomic nuclei is estimated to exceed the size of a nucleus.
Therefore it is believed that there is essentially no room for spatially inhomogeneous excitations of pairing modes.
The second one is related to the fact that a nucleus has a finite number of nucleons and consequently its description requires projecting out (from the BCS wave function) the component having well defined particle number. 
However, once this operation is performed the information about the phase is lost, at least to some extent.
Still these arguments cannot rule out the existence of observable consequences of pairing field dynamics and they may only serve as a warning sign when extracting
quantitative predictions based on the mean-field treatment of pairing correlations.

There have been only a limited number of attempts to investigate features of nuclear pairing that could serve as a proof that it can be treated as a superfluid phase.
They all focused on probing the degree of freedom of pairing field related to the spatial modulation of its phase factor.
Some of the first examples of such studies were triggered by the discovery of the Josephson effect in condensed matter systems and were devoted to searching for its analog in atomic nuclei~\cite{goldanski,dietrich1,dietrich2,dietrich3}. 
In the case of nuclear systems it was investigated in collisions of two nuclei, where the relative phase difference was predicted to induce enhanced transfer of neutrons.
These studies, however, did not lead to conclusive results~\cite{mermaz1,mermaz2,mermaz3,sugiyama}.
Recently, a new light has been shed on the problem, after realizing that the collision of two nuclei creates rather an AC Josephson junction which induces oscillations of nucleons and thus may lead to photons emission of certain energy~\cite{potel21, magierski21}.
Surprisingly enough the predicted gamma energies have been found to be in agreement with experiment \cite{potel21,montanari14,montanari16}.
All these studies are related to sub-barrier collisions and therefore tunneling plays a crucial role in their description.

Another type of studies have been performed above the barrier, where two nuclei merge having two different pairing phases~\cite{hashimoto16, magierski17}.
From physical point of view the situation is different. 
The collision above the barrier requires more energy and the process is not nearly adiabatic as in the case of sub-barrier reactions. 
Consequently it leads to creation of solitonic excitation, which was described in Ref.~\cite{magierski17}.
An analog of such collision is routinely realized with ultracold atomic clouds, where the non adiabatic character of dynamics is reflected in creation of the solitonic excitation which subsequently decays~\cite{MIT1,MIT2,wlazlowski18}.
In nuclear collisions it is predicted to enhance the effective barrier between colliding nuclei~\cite{magierski17}.
The extra energy needed to overcome the additional repulsion depending on phase difference between colliding nuclei resembles extra-push energy (see Refs.~\cite{swiatecki1982,swiatecki1982a,donangelo1986,Washiyama2015}) although has a completely different physical origin.

This paper represents the continuation of studies reported in Ref.~\cite{magierski17} and the first case of describing the collision of superfluid medium-mass nuclei with a realistic nuclear energy density functional. 
To allow for comparison to previous results obtained in Ref.~\cite{magierski17} the collisions of $^{90}$Zr+$^{90}$Zr and $^{96}$Zr+$^{96}$Zr were 
chosen.
Moreover we have investigated the influence of pairing dynamics on the evolution of deformation of composite system. 
Finally, a qualitatively new effect is described which leads to dynamically induced $U(1)$ symmetry breaking leading to the enhancement of pairing correlations after collision.

\ssec{Theoretical framework}
The theoretical method is based on TDDFT extended to superfluid systems with local density approximation, TDSLDA (time-dependent superfluid local density approximation).
The approach has been described in Refs.~\cite{Bulgac:2002uq,LNP2012,ARNPS2013,Mag2016,LISE} and its particular realization for nuclear dynamics is given in Ref.~\cite{Mag2016}.
It amounts to solving the equations for both protons and neutrons:
\begin{widetext}
\begin{eqnarray} \label{tdslda}
i\hbar\frac{\partial}{\partial t} 
\left  ( \begin{array} {c}
  U_{\uparrow   n}({\bf r},t)\\  
  U_{\downarrow n}({\bf r},t)\\
  V_{\uparrow   n}({\bf r},t)\\ 
  V_{\downarrow n}({\bf r},t)\\
\end{array} \right ) =
\left ( \begin{array}{cccc}
h_{\uparrow \uparrow }({\bf r},t)& h_{\uparrow \downarrow }({\bf r},t)  & 0 & \Delta_{\uparrow\downarrow}({\bf r},t)\\
h_{\downarrow \uparrow }({\bf r},t)& h_{\downarrow \downarrow }({\bf r},t) & \Delta_{\downarrow\uparrow}({\bf r},t) & 0 \\
0& \Delta^{*}_{\downarrow\uparrow}({\bf r},t) & -h^{*}_{\uparrow\uparrow} ({\bf r},t)& -h^{*}_{\uparrow \downarrow }({\bf r},t) \\
\Delta^{*}_{\uparrow\downarrow}({\bf r},t)& 0 & -h^{*}_{\downarrow \uparrow } ({\bf r},t) & -h^{*}_{\downarrow \downarrow }({\bf r},t) \\
\end{array} \right )  
\left  ( \begin{array} {c}
  U_{\uparrow   n}({\bf r},t)\\  
  U_{\downarrow n}({\bf r},t)\\
  V_{\uparrow   n}({\bf r},t)\\ 
  V_{\downarrow n}({\bf r},t)\\
\end{array} \right ),
\end{eqnarray}
\end{widetext}
where $h_{ij}({\bf r},t)$ is the mean-field term which is obtained from the Skyrme SkM* functional \cite{SkMstar}.
The pairing field $\Delta_{\uparrow\downarrow}({\bf r},t) = -\Delta_{\downarrow\uparrow}({\bf r},t) = g({\bf r})\sum_{n}V^{*}_{\uparrow n}({\bf r},t)U_{\downarrow n}({\bf r},t)$ will be denoted from now on as $\Delta({\bf r},t)$.
The running coupling constant $g({\bf r})$ depends on the local momentum cutoff according to the prescription given in Refs.~\cite{BY:2002fk,Bulgac:2002uq,Mag2016,LISE}. 
Its explicit form reads:
\begin{widetext}
\begin{subequations}
\begin{eqnarray}
\frac{1}{g({\bf r})} &=& \frac{1}{g_{0}}-\frac{m^{*}({\bf r})k_{c}({\bf r})}{2\pi^{2}\hbar^{2}}
\left ( 1 - \frac{k_{F}({\bf r})}{2k_{c}({\bf r})} \log \left [ \frac{k_{c}({\bf r})+k_{F}({\bf r})}{k_{c}({\bf r})-k_{F}({\bf r})} \right ] \right ), 
\textrm{ for }k_{F}^{2}({\bf r})\ge 0  \\
\frac{1}{g({\bf r})} &=& \frac{1}{g_{0}}-\frac{m^{*}({\bf r})k_{c}({\bf r})}{2\pi^{2}\hbar^{2}}
\left ( 1 + \frac{k_{F}({\bf r})}{k_{c}({\bf r})} \arctan \left[ \frac{|k_{F}({\bf r})|}{k_{c}({\bf r})} \right ] \right ), 
\textrm{ for }k_{F}^{2}({\bf r})< 0  \\
E_{c} &=& \frac{\hbar^{2}k_{c}^{2}({\bf r})}{2m} + \mathcal{U}({\bf r}) - \mu, \label{eq:rkcdef}\\
\mu  &=& \frac{\hbar^{2}k_{F}^{2}({\bf r})}{2m} + \mathcal{U}({\bf r}),\label{eq:rkFdef}
\end{eqnarray}
\label{eq:renorm}
\end{subequations}
\end{widetext}
where the above equations are defined separately for neutrons and protons. 
The quantity $g_0$ denotes bare coupling constant. 
Equations.~(\ref{eq:rkcdef}) and (\ref{eq:rkFdef}) define local momentum cutoff $\hbar k_c({\bf r})$ via energy cut-off $E_c$ and local Fermi momentum $\hbar k_F({\bf r})$, respectively. 
The mean-field potential generated by nucleons with effective mass $m^*({\bf r})$ is denoted by $\mathcal{U}({\bf r})$, and chemical potential $\mu$ is used to control the average particle number. 

Equations (\ref{tdslda}) are solved on the spatial lattice with
20\,$\times$\,20\,$\times$\,64 grid points in the case of $^{90}$Zr+$^{90}$Zr collisions and 24\,$\times$\,24\,$\times$\,64 for $^{96}$Zr+$^{96}$Zr collisions, with 1.25\,fm grid spacing in each direction. 
Periodic boundary conditions are applied to the box. 
The numerical evolution is performed using the fifth-order predictor-modifier-corrector Adam-Bashforth-Milne method.

In order to obtain the initial condition for time evolution, which consists of two nuclei separated by 40\,fm we have used the conjugate orthogonal conjugate gradient (COCG) method~\cite{COCG}.
An external potential was used to counteract the Coulomb repulsion and to
keep the centers of mass of the two nuclei at rest. 
The same technique was used in Refs.~\cite{magierski17,barton20}.
The COCG method makes use of Green's functions to obtain various densities, without diagonalizing the Hamiltonian. 
After obtaining convergent self-consistent densities, they are subsequently inputted into a code which generates the wavefunctions by diagonalizing the Hamiltonian. 

The momentum cutoff used in these studies corresponds to $E_c=100$ MeV energy cutoff. 
The energy cutoff defines the number of $U$ and $V$ components which are evolved and it has an impact on computational complexity. 
For static calculations, it can be shown analytically that the result does not depend on the energy cutoff, if it is set to a sufficiently large value \cite{BY:2002fk,Bulgac:2002uq}. 
On the other hand, there is a relation between the value of the cutoff and the length of the numerically accurate (stable) time evolution which can be traced back to the properties of the Bogoliubov transformation~\cite{Mag2016, Mag18}.
The larger cutoff allows for longer trajectories.
For collisions between nuclei where the expected length of trajectory does not exceed $4000$\,fm/$c$ the energy cutoff $E_{c}=100$ MeV is sufficient~\cite{grineviciute18}.

Numerical stability is best reflected in energy and particle number conservation
during the evolution. 
The deviation of the energy from its initial value ~$\Delta E_{tot} = |E_{tot}(t) - E_{tot}(t=0)|/|E_{tot}(t=0)|$ did not exceed $8\times 10^{-4}$, which corresponds to energy variations less than $1.3$ MeV for each run.
In most simulations relative energy change was about $\Delta E_{tot} \lesssim 4\times 10^{-4}$.
The quantity describing change of fluctuation the average particle number reads: ~$\Delta N_{q} = |N_{q}(t) - N_{q}(t=0)|$, where $q=n,p$ denotes neutrons and protons, respectively. 
For all runs $\Delta N_{q} < 10^{-4}$ for both protons and  neutrons.
The time step $\Delta t$ in calculations is related to the lattice spacing $a$: $\Delta t < \frac{2 m a^{2}}{\hbar \pi^{2}}$. 
For the lattice spacing $a=1.25$~fm we chose $\Delta t = 0.119956$ fm/cm, which ensures numerical stability within the evolution time considered in this paper.

\begin{figure}[t]
\centering
\includegraphics[width=\columnwidth]{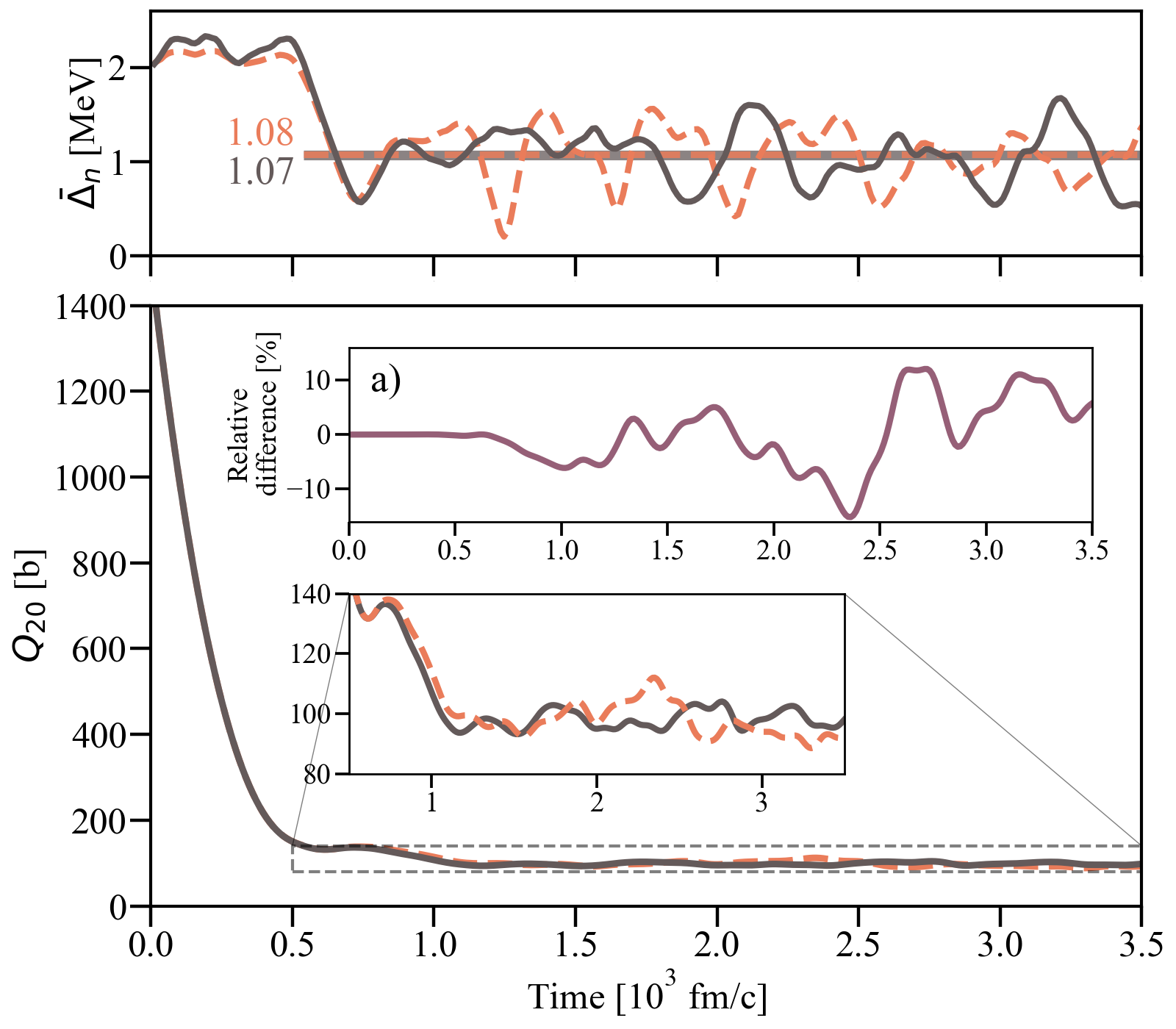}
\caption{Evolution of neutron pairing gap $\overline\Delta_n$ (top panel) and quadrupole moment $Q_{20}$ (bottom panel) during the head-on collision of $^{96}$Zr+$^{96}$Zr at center-of-mass energy $185$~MeV. 
Dashed (orange) lines show the evolution obtained using the regularization scheme~(\ref{eq:renorm}) with $g_{0}=280$ MeV fm$^{3}$. 
Solid (gray) lines show the evolution obtained with fixed coupling constant $g_0^\prime=190 $ MeV fm$^{3}$. 
In the top panel the average pairing gap after the collision is presented 
(horizontal line).
In the inset (a) of the bottom panel the evolution of the relative difference of the quadrupole moment is shown.
}
\label{fig:sm-response}
\end{figure}

The main results, presented in next sections, were executed with the regularization prescription given by the relations~(\ref{eq:renorm}). 
However, we also confirmed stability of the results with respect to the regularization scheme. 
Namely, we performed test calculations with fixed coupling constant. 
We set $\frac{1}{g({\bf r})} = \frac{1}{g_0^\prime}$, where $g_0^\prime$ was  chosen separately for protons and neutrons to match the average pairing gap (within $0.02$ MeV accuracy) obtained in calculations using the regularization~(\ref{eq:renorm}).
Other simulation parameters remained unchanged.
The difference of the total energy between the two approaches did not exceed $0.5$ MeV within the time interval of  simulations.
In Fig.~\ref{fig:sm-response} we present the time evolution of neutron average pairing gap $\overline\Delta_n$ (\ref{avdel}) and the quadrupole moment $Q_{20}$ (\ref{quad}) as obtained within two schemes (see also movies in the Supplemental Material \cite{SM}).
For this test we consider a head-on collision of $^{96}$Zr+$^{96}$Zr at the center-of-mass energy $185$ MeV. 
As shown in the bottom panel of Fig.~\ref{fig:sm-response}, the shape evolution of the system, measured by its quadrupole moment $Q_{20}$, evolves, in both cases, in the same way. 
The relative difference of quadrupole moments oscillates around zero with maximum deviation of approximately $10\%$. 
The differences in the evolution of the pairing gap are more pronounced; however, closer inspection shows that in both cases gross properties remain unchanged. 
Namely, after the collision, the pairing gap oscillates around the same average value and both the amplitude of oscillations and their period remain the same.
The executed tests confirmed that overall system evolution, and thus quantities discussed in this paper, are not sensitive to the regularization scheme.

\begin{figure}[t]
   \begin{center}
   \includegraphics[width=0.98\columnwidth]{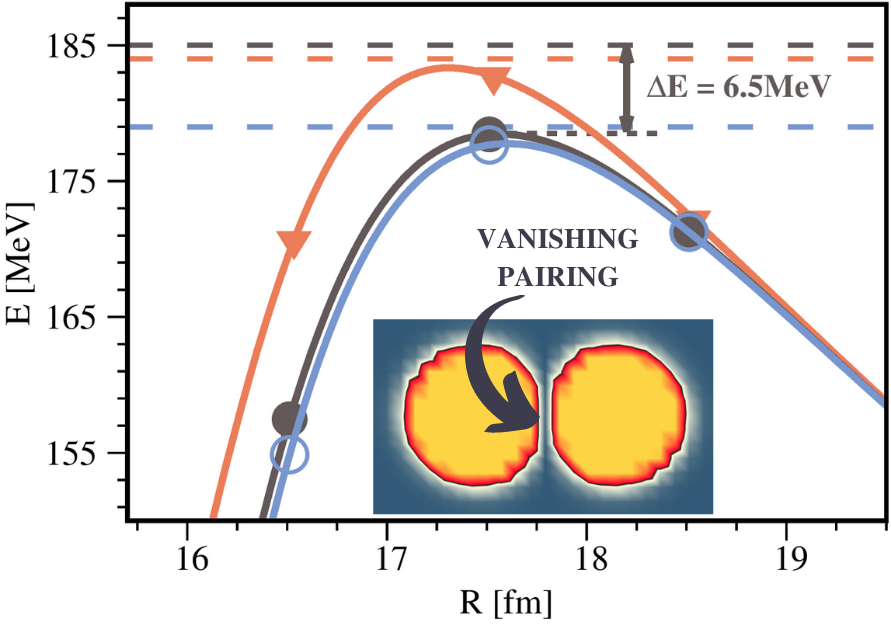}
   \end{center}\vspace{-3mm}
   \caption{
   Nucleus-nucleus potential calculated in the frozen density approximation with inclusion of the pairing field.
   The barrier was calculated for the cases of $^{90}$Zr+$^{90}$Zr (solid line with filled triangles) and $^{96}$Zr+$^{96}$Zr with relative phase of pairing fields $\Delta\phi=0$ (solid line with open circles) and $\Delta\phi=\pi$ (solid line with filled circles).
    Horizontal dashed lines indicate the minimum energy for capture, $\Ethresh$ (see Table \ref{tab:ethresh}).
    In the inset we show a snapshot of the magnitude of the neutron pairing field from collisions with $\Delta\phi=\pi$, where a solitonic excitation is formed between nuclei.
   }
   \label{fig:frozendensity}
\end{figure}

\ssec{Solitonic excitation and its decay}
The outcome of a collision of two nuclei described within TDSLDA can be naturally divided into two regimes: capture or reseparation.
Due to the fact that TDSLDA does not take into account sub-barrier tunneling of the many-body wavefunction the separation between the two cases can be well defined by a certain threshold energy $\Ethresh$ (we consider here head-on collisions only). 
In TDSLDA [where $U(1)$ symmetry is broken] superfluidity is represented by the complex pairing field and thus allows for a freedom of setting the relative phase factors of pairing fields of colliding nuclei. 
As a consequence the threshold energy will be, in principle, a function of the relative gauge angle: ${\Ethresh(\Delta\phi)}$.
This dependence was investigated in Ref.~\cite{hashimoto16} for light systems and in Ref.~\cite{magierski17} for medium and heavy nuclei and has been attributed to the appearance of solitonic excitation during the collision~\cite{magierski17}.
It was shown that the dependence of the threshold energy on the relative gauge angle appears as a result of creation of a junction between colliding nuclei where the pairing field exhibits strong spatial variation. 
Consequently, part of energy is stored in the junction which can be estimated from the Ginzburg-Landau model to be proportional to 
\begin{equation}
\Delta E_{s} \propto \sin^2 \left(\frac{\Delta\phi}{2} \right)
\label{jj}
\end{equation}
This energy gives rise to an effective increase of the threshold energy for capture.
It also implies that the variation of $\Ethresh$ is of dynamic character, which cannot be explained by the static calculations, even with inclusion of the pairing field.
In order to support this statement, in Fig.~\ref{fig:frozendensity} we present a comparison between static barriers obtained in the frozen density approximation and the TDSLDA threshold energies, which are shown for collisions $^{96}$Zr+$^{96}$Zr and $^{90}$Zr+$^{90}$Zr. 
In the frozen density approximation the dynamical effects during the collision are neglected and the density of each fragment is fixed to be its ground-state one.
Therefore it slightly overestimates the height of the barrier and, compared to the density-constrained TDHF, one expects that for the presented collisions 
the difference reads 2-3\,MeV \cite{Washiyama2015}.
In the frozen density approximation the contribution from the pairing fields 
were also taken into account. 
Namely, the pairing contribution to the total energy was generated by the pairing field:
\begin{equation}
\Delta_{q,\text{tot}}({\bf r})=\Delta_{q,1}({{\bf r} - R/2}) + \Delta_{q,2}({{\bf r} + R/2}),
\end{equation}
where $q=n,p$ denote neutron and proton contributions and $1,2$ specify the initial nuclei. 
The parameter $R$ denotes the distance between centers of mass of two nuclei.
Since we consider collisions of two identical nuclei, $\Delta_{q, 1}({\bf r }) = \exp( i\Delta\phi ) \Delta_{q, 2}({\bf r})$.
Consequently the static barriers differ for $\Delta\phi=0$ and $\Delta\phi=\pi$ cases, since in the latter case the overlapping pairing fields of two colliding nuclei cancel out.
The dashed lines correspond to the threshold energy $\Ethresh$ obtained within TDSLDA, which is the energy at which merging occurs.
This energy has been determined with 1-MeV accuracy, and the condition for merging implies that the two nuclei form a composite system and do not separate within a time interval of 4000\,fm/$c$. 
It has to be emphasized that the transition between the two regimes is fairly sharp, so the changes of this time interval by $\pm$2000\,fm/$c$ do not affect the extracted value of the threshold energy.

In Fig.~\ref{fig:frozendensity} one may notice that the location of threshold energy agrees relatively well with the maximum of the barrier calculated in the frozen density approximation when there is no phase difference.
The situation is dramatically different in the case of phase difference equal to $\pi$.
In this case the barrier height in the frozen density approximation is about 1-2\,MeV higher due to the cancellation of pairing fields belonging to colliding nuclei. 
However, this increase of the barrier due to static pairing contribution does not reproduce the threshold energy in this case.
The difference between the static barrier and the actual threshold energy obtained in time-dependent calculations represents the contribution coming from formation of the solitonic excitation between colliding nuclei.
It therefore indicates that the effect is of dynamic origin, which cannot be grasped in the static calculations.
As was conjectured in  Ref.~\cite{magierski17}, the contribution comes from the term $\int|\nabla\Delta({\bf r})|^{2}d^{3}{\bf r}$, present in the Ginzburg-Landau model, which produces additional barrier depending on the relative difference of gauge angles according to Eq.~(\ref{jj}).
The coefficient in front of Eq.~(\ref{jj}) may depend on the pairing strength; however, it is not {\em a priori} clear to what extent pairing magnitude can affect $\Ethresh$.
In order to address this question, we list in Table~\ref{tab:ethresh} the values of $\Delta E_{s}$ are listed, which measure the differences between threshold energies corresponding to $0$ and $\pi$ phase differences (see also Fig.~\ref{fig:frozendensity}).
In the case of $^{96}$Zr+$^{96}$Zr we have performed collisions for three sets of the average pairing gaps, defined as
\begin{equation} \label{avdel}
\overline{\Delta}_q = \frac{1}{N_q}\int d^{3}r |\Delta_{q}({\bf r})|\rho_q({\bf r}),
\end{equation}
where $q=n,p$ and $N_q$ denotes the total number of neutrons or protons.
Namely, we varied the average of neutron pairing gaps, which has a dominant contribution to the effect, while keeping the proton pairing gap approximately constant.
It can be noticed that the dependence is not linear and the effect is reduced as compared to the values reported in Ref.~\cite{magierski17}, where Fayans energy density functional FaNDF$^0$ without spin-orbit term was employed. 
It is, however, consistent with estimations of the impact of superfluidity on the fusion barrier extracted from experimental data in Ref.~\cite{scamps18}, where the effect is concluded to be weaker.

\begin{table}[t]
\centering
\caption{The minimum energies needed for capture in $^{90}$Zr+$^{90}$Zr and $^{96}$Zr+$^{96}$Zr for the case of $\Delta\phi=0$ [$\Ethresh(0)$]  and $\Delta\phi=\pi$ [$\Ethresh(\pi)$]. 
The energy difference between the two cases is shown in the last column. 
The average pairing gap $\overline{\Delta}_{i}$ is defined by Eq.~(\ref{avdel}).}
\begin{tabular}{c|c||c|c|c} 
\hline\hline
                            &    $\overline{\Delta}_q$ (MeV) & $\Ethresh(0)$ (MeV) & $\Ethresh(\pi)$ (MeV) & $\Delta E_s$  \\ 
\hline
\multirow{2}{*}{$^{90}$Zr } & $\overline{\Delta}_n = 0.00$  & \multirow{2}{*}{184} & \multirow{2}{*}{184}  & \multirow{2}{*}{0}  \\ 
                            & $\overline{\Delta}_p = 0.09$  &                      &                       &                     \\ \hline
\multirow{6}{*}{$^{96}$Zr } & $\overline{\Delta}_n = 1.98$  & \multirow{2}{*}{179} & \multirow{2}{*}{185}  & \multirow{2}{*}{6}  \\ 
                            & $\overline{\Delta}_p = 0.32$  &                      &                       &                     \\ \cline{2-5}
                            & $\overline{\Delta}_n = 2.44$  & \multirow{2}{*}{178} & \multirow{2}{*}{187}  & \multirow{2}{*}{9}  \\ 
                            & $\overline{\Delta}_p = 0.33$  &                      &                       &                     \\ \cline{2-5}
                            & $\overline{\Delta}_n = 2.94$  & \multirow{2}{*}{178} & \multirow{2}{*}{187}  & \multirow{2}{*}{9}  \\ 
                            & $\overline{\Delta}_p = 0.34$  &                      &                       &                     \\
\hline\hline
\end{tabular}
\label{tab:ethresh}
\end{table}

Since solitonic excitation leads to an effective barrier increase, it also contributes to increased repulsion between nuclei.
This effect, although obvious, when one considers the Ginzburg-Landau approximation may sound strange when one thinks about superfluidity as a lubricant facilitating nuclear large-amplitude collective motion.
In order to quantify this effect we considered the total excitation energy (TXE) of the fragments at sub-barrier energies $\Ecm < \Ethresh$. 
In Table~\ref{tab:txe} the total excitation energies are shown for collision energies just below $\Ethresh(0)$.
They were evaluated according to the prescription $\text{TXE} = \Ecm - \text{TKE}$, where TKE denotes the asymptotic total kinetic energy of the outgoing fragments.
It can be seen that for the same energy of colliding nuclei the case corresponding to $\Delta\phi=\pi$ leads to significantly lower TXE. 
Moreover, with increasing magnitude of the pairing field this quantity is decreasing quite rapidly, contrary to the $\Delta\phi=0$ case where TXE decreases by about 10\% only.
The sensitivity of TXE on pairing magnitude for $\Delta\phi=\pi$ comes from the fact that increasing magnitude of pairing field leads to stronger repulsion between fragments. 
Consequently for the same sub-barrier energy the nuclei colliding with $\Delta\phi=\pi$ experience significantly smaller overlap between densities at the distance of the closest approach, than those colliding at $\Delta\phi=0$.
This is another confirmation of the impact of solitonic excitation on nuclear dynamics which effectively acts as a ``spring'' between colliding nuclei, absorbing part of the energy and thus giving rise to additional repulsion.
The absorbed energy can be either released in the form of increased TKE (for $\Ecm < \Ethresh$) or dissipated to other degrees of freedom (for $\Ecm > \Ethresh$).

\begin{table}
\centering
\caption{Total excitation energies (TXEs) in $^{96}$Zr+$^{96}$Zr
at c.m. energies $\Ecm$ just below the threshold for capture (see Table~\ref{tab:ethresh})
with $\Delta\phi=0$ [TXE(0)] and $\Delta\phi=\pi$ [TXE($\pi$)]. 
The average pairing gap $\overline{\Delta}_{i}$ is defined by Eq.~(\ref{avdel}).}
\begin{tabular*}{\columnwidth}{@{\extracolsep{\fill}}c|c||c|c|c}
\hline\hline
 & \multirow{2}{*}{$\overline{\Delta}_q$ (MeV)} & \multirow{2}{*}{$\Ecm$ (MeV)} & \multicolumn{2}{c}{TXE (MeV)} \\ \cline{4-5}
                            & & & $\Delta\phi=0$ & $\Delta\phi=\pi$  \\ 
\hline
\multirow{6}{*}{$^{96}$Zr } & $\overline{\Delta}_n = 1.98$ & \multirow{2}{*}{178} & \multirow{2}{*}{37}  & \multirow{2}{*}{25}  \\ 
                            & $\overline{\Delta}_p = 0.32$ &                      &                       &                     \\ \cline{2-5}
                            & $\overline{\Delta}_n = 2.44$ & \multirow{2}{*}{177} & \multirow{2}{*}{34}  & \multirow{2}{*}{10}  \\ 
                            & $\overline{\Delta}_p = 0.33$ &                      &                       &                     \\ \cline{2-5}
                            & $\overline{\Delta}_n = 2.94$ & \multirow{2}{*}{177} & \multirow{2}{*}{34}  & \multirow{2}{*}{8}  \\ 
                            & $\overline{\Delta}_p = 0.34$ &                      &                       &                     \\
\hline\hline
\end{tabular*}
\label{tab:txe}
\end{table}

\begin{figure}[t]
\begin{center}
\includegraphics[width=0.98\columnwidth]{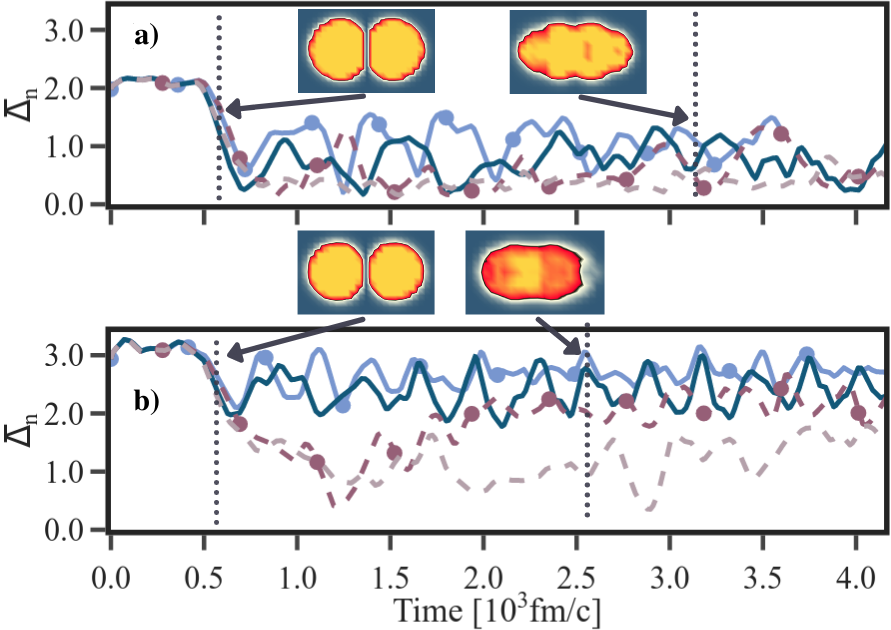}
 \end{center}\vspace{-3mm}
\caption{Neutron average pairing gap $\overline{\Delta}_{n}$ as a function of time for $^{96}$Zr+$^{96}$Zr for two initial magnitudes of neutron pairing [(a) 2\,MeV, (b) 3\,MeV].
Solid lines represent the results obtained for $\Delta\phi=0$, and dashed lines represent those for $\Delta\phi=\pi$.
The results for energies $\Ecm=185$\,MeV in (a) and $\Ecm=187$\,MeV in (b) are denoted by solid circles. 
The curves without symbols correspond to energy $\Ecm=193$\,MeV. 
Insets show snapshots of the magnitude of neutron pairing field for collisions at lower energies ($185$ and $187$\,MeV) with $\Delta\phi=\pi$ at two selected times indicated by arrows.
}
\label{fig:pairn}
\end{figure}

In the case of $\Ecm > \Ethresh$ the energy absorbed by the solitonic excitation is dissipated, since it is immersed in the nuclear environment and thus is coupled strongly to other collective and single-particle degrees of freedom. 
Consequently the soliton will eventually decay and its energy will be distributed among other degrees of freedom. 
Its decay is correlated with the behavior of the average value of the pairing gap.
The dynamics of the pairing field during the collision process is presented in Fig.~\ref{fig:pairn}, where the evolution of the average neutron pairing gap $\overline{\Delta}_{n}$ is shown (see also movies in the Supplemental Material \cite{SM}).
At the moment of collision the average pairing gap drops considerably due to breaking of Cooper pairs. 
The peculiarity of the superfluid dynamics is seen the best in the case of the strongest pairing.
The initial merging of two nuclei leads to significant decrease of pairing magnitude in the first stage of the collision process, which particularly visible in the case of $\Delta\phi =\pi$, and then is followed by restoration of the pairing strength. 
The timescale of the restoration process is of the order of 2000-6000\,fm/$c$ and depends on the collision energy. 
The process of the pairing restoration after the initial decrease due to the collision is also seen for weaker pairing strengths although it is less pronounced.
It turns out that the decay process of the solitonic excitation occurs approximately at times when the restoration process of pairing starts.
In the simulations presented here it corresponds to times 1500-2000\,fm/$c$ after the collision. 
The insets of Fig.~\ref{fig:pairn} show examples of neutron pairing field distribution after merging at times when solitonic excitation is still presents and when it is gone.

\ssec{Pairing dynamics and evolution of deformation of a composite system}
TDSLDA predicts yet another qualitative difference in the evolution of deformation of a composite system in the presence of pairing.
It has to be emphasized that the pairing dynamics in TDSLDA is independent of
density fluctuations although both are dynamically coupled. 
This is contrary to, e.g., the TDHF+BCS approach~\cite{scamps12}, where the pairing field is constructed as already coupled to the nucleon density. 
This has serious consequences as a variety of excitation modes of a superfluid system cannot be described within the latter approach (see Ref.~\cite{Mag18} for details).

\begin{figure}[t]
\centering
\includegraphics[width=1.02\columnwidth]{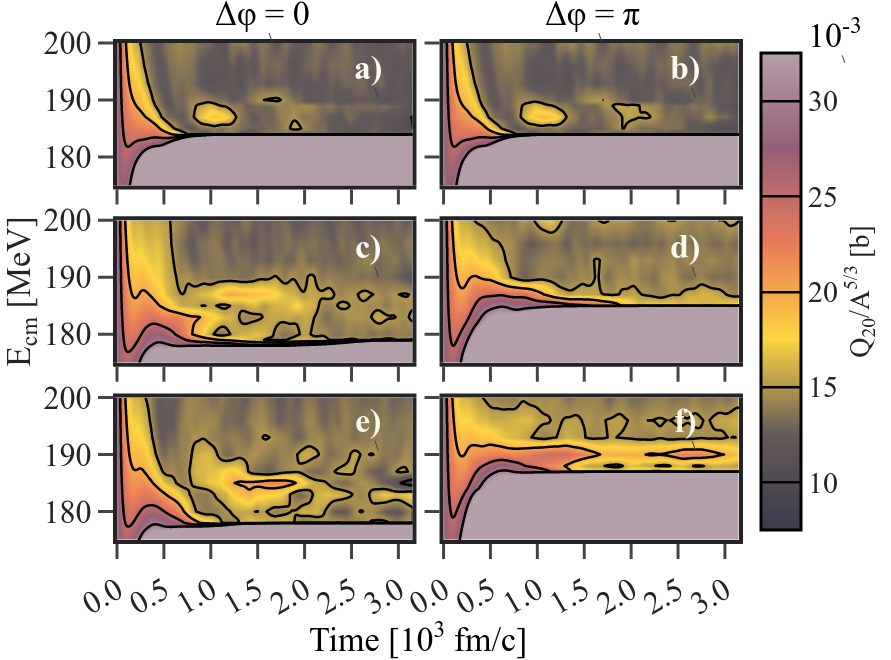}
\caption{Rescaled quadrupole moment $Q_{20}/A^{5/3}$ ($A$ is the total nucleon number of the system) of colliding nuclei as a function of $\Ecm$ and time. 
Panels (a) and (b) correspond to reaction $^{90}$Zr+$^{90}$Zr ($\overline{\Delta}_n=0.002$\,MeV, $\overline{\Delta}_p=0.09$\,MeV). 
Other panels correspond to $^{96}$Zr+$^{96}$Zr with different pairing strengths: $\overline{\Delta}_n=1.98$\,MeV, $\overline{\Delta}_p=0.32$\,MeV [panels.~\ref{fig:q20map}(c) and (d)] and $\overline{\Delta}_n=2.94$\,MeV, $\overline{\Delta}_p=0.34$\,MeV [panels.~\ref{fig:q20map}(e) and (f)].
The case of $\Delta\phi=0$ is shown in panels: (a), (c), (e) whereas the case of $\Delta\phi=\pi$ is shown in panels: (b), (d), (f).
Time t=0 is chosen to be when the quadrupole moment goes down and reaches value $Q_{20}/A^{5/3} = 0.03$b.
It corresponds to approximately 100 fm/c before neck formation.
The value $Q_{20}/A^{5/3} = 0.03$b sets also the upper limit of deformation shown in the figure.
}
\label{fig:q20map}
\end{figure}

Evolution of deformation of the composite system after collision is shown in Fig.~\ref{fig:q20map}, where the quadrupole moment is plotted as a function of time and collision energies (see also movies in the Supplemental Material\cite{SM}).
Since the head-on collisions, considered here, possess a symmetry axis, the elongation can be measured conveniently using quadrupole moment with respect to the
center of mass of two nuclei:
\begin{equation}\label{quad}
Q_{20} = \int d^{3}{\bf r} (3 z^{2} - r^{2} ) \rho({\bf r}) ,
\end{equation}
where the $z$ axis is assumed to be the symmetry axis.
From the figure we find that the evaluations of deformation in collisions of neutron magic nuclei $^{90}$Zr+$^{90}$Zr and in reaction $^{96}$Zr+$^{96}$Zr are qualitatively different.
In the former case [Figs.~\ref{fig:q20map}(a) and \ref{fig:q20map}(b)], which correspond to vanishingly small pairing (see Table \ref{tab:ethresh}) the system becomes more compact right after collision.
In contrast, the evolution of deformation of $^{96}$Zr+$^{96}$Zr system is more complex.
The quadrupole moment reveals variations as a function of time, which are attributed to enhanced susceptibility towards shape changes caused by pairing correlations.
Moreover, the evolution of the deformation towards compact shape is visibly delayed in the presence of superfluidity.
This effect is particularly pronounced at energies close to the barrier in the case of $\Delta\phi=\pi$.

\begin{figure}[t]
\centering
\includegraphics[width=1.02\columnwidth]{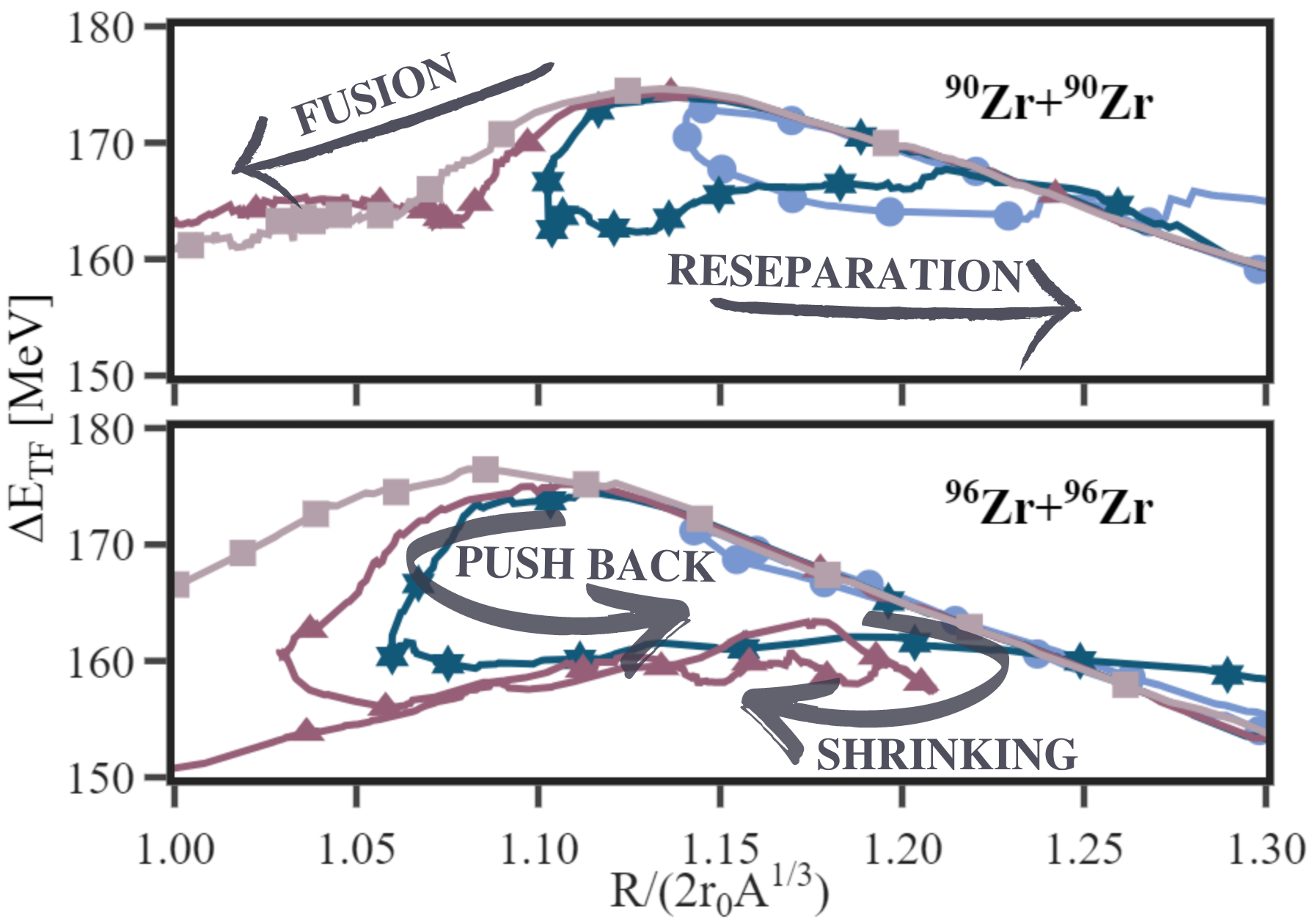}
\caption{Energy of colliding nuclei in the Thomas-Fermi approximation (see text for details) as a function of distance between their centers of mass ($r_{0}=1.2$ fm)
for four $\Ecm$ values. 
$\Delta E_{TF}$ denotes the energy with respect to the configuration with two nuclei at infinite distance. 
The upper panel shows the case of $^{90}$Zr+$^{90}$Zr ($\Delta\phi=0$, $\overline{\Delta}_{n}=0.002$ MeV, $\overline{\Delta}_{p}=0.09$ MeV).
The lower panel presents the case of $^{96}$Zr+$^{96}$Zr ($\Delta\phi=\pi$, 
$\overline{\Delta}_{n}=2.94$ MeV, $\overline{\Delta}_{p}=0.34$ MeV).}
\label{fig:thomasfermi}
\end{figure}

The delayed evolution of the deformation in the presence of pairing indicates another feature associated with neck formation.
Namely, the system, after developing a neck, is considerably {\em pushed back}, increasing its deformation, until it shrinks again (at smaller energy this effect leads to reseparation).
The effect can be seen in Fig.~\ref{fig:thomasfermi}, where the
total energy of the system in the Thomas-Fermi (TF) approximation is plotted.
The TF energy $E_\text{TF}$ is defined as
\begin{equation}
\begin{split}
E_\text{TF}& = E_\text{Coul} + \int d^{3}{\bf r} \Bigg [ \sum_{q=n,p}\frac{\hbar^{2}}{2m_q}\tau^\text{TF}_{q} \\
+&  \sum_{t=0,1} \Bigg ( C^{\rho}_{t}\rho_{t}^{2} 
+ C^{\Delta\rho}_{t}\rho_{t}\nabla^{2}\rho_{t} 
       + C^{\tau}_{t}\rho_{t}\tau^\text{TF}_{t}  + C^{\gamma}_{t}\rho_{t}^{2}\rho^{\gamma}_{0} \Bigg ) \Bigg ] 
       \label{eq:tf_energy}
\end{split}
\end{equation}
where isoscalar and isovector nucleon densities, $\rho_{0}=\rho_{n}+\rho_{p}$ and $\rho_{1}=\rho_{n}-\rho_{p}$, respectively, are obtained from TDSLDA calculations and the kinetic energy distribution is taken as  $\tau^\text{TF}_q = \frac{3}{5}(3\pi^{2})^{2/3} \rho_q^{5/3} + \frac{1}{36}\frac{\left ( \nabla\rho_{q} \right )^{2}}{\rho_{q}} + \frac{1}{3}\nabla^{2}\rho_{q} $, where the second-order correction to standard the TF approximation was included.
Isoscalar and isovector components $\tau^\text{TF}_{t}$ are defined similarly to densities $\rho_{t}$.
The quantity $E_\text{TF}$ includes the static part of the energy of colliding nuclei (excluding collective flow energy), allowing us to follow the energy changes due to evolution of the deformation.
In Fig.~\ref{fig:thomasfermi} this quantity is plotted as a function of the relative distance.
The formation of the neck is visible as a rapid drop of initially increasing energy.
Subsequently, in the case of $^{90}$Zr+$^{90}$Zr, the trajectory either goes back towards reseparation or leads to shrinking and developing a more compact configuration.
Contrarily, in the case of $^{96}$Zr+$^{96}$Zr there is an energy window in which another regime is possible. 
Namely, the merged fragments are pushed back, increasing the quadrupole moment. Subsequently, however, this process does not lead to reseparation since the system stops and returns to forming a compact system.
It indicates that the presence of pairing alters significantly the process
of neck formation.

\ssec{Enhancement of pairing correlations as a result of collision}
Pairing fields of two colliding nuclei undergo a significant weakening as a result of kinetic energy transfer to internal degrees of freedom, leading to Cooper pair breaking.
Surprisingly the inverse process, i.e., the enhancement of pairing correlations due to collision, is possible as well.
Suppose that colliding nuclei have a relatively small pairing field, which is due to their shell structure.
The collision of such nuclei may increase the magnitude of the pairing field of the composite system.
The mechanism of such behavior is simple. 
The initial pairing strength is dictated by the product of the pairing coupling constant $g$ and density of states at the Fermi surface.
Namely, in the weak coupling limit the relation reads
\begin{equation}
\Delta \propto \exp\biggl[-\frac{2}{ g N(\epsilon_{F}) }\biggr],
\end{equation}
where $N(\epsilon_{F})$ is the density of states at the Fermi surface.
Due to collision a composite system is formed where the relation $N_{c}(\epsilon_{F})\gg N(\epsilon_{F})$ can hold.
Consequently the effect will be similar to increasing artificially the pairing coupling constant and thus causing instability and leading to the symmetry breaking.
In this case, however, this is due to the change of geometry of the system, which affects the density of states.

One has to keep in mind, however, that these are static considerations which correspond to the limit in which we have a system in equilibrium at the initial time, which subsequently evolves under external perturbation (in this case caused by collision) to another equilibrium state, characterized by a different density of states.
This picture is oversimplified in the situation of colliding nuclei.
The created composite system will eventually equilibrate but at much longer time-scale. 
At the short timescale which we consider in this paper, the composite system is still far from equilibrium and the concept of density of states does not have a clear meaning. 
Moreover there is a significant amount of energy deposited in the system which weakens pairing in the composite system.
Therefore, although the presented arguments may remain valid, there are various factors related to nuclear collision dynamics which may change the actual evolution of the pairing field.

\begin{figure}[t]
\centering
\includegraphics[width=0.93\columnwidth]{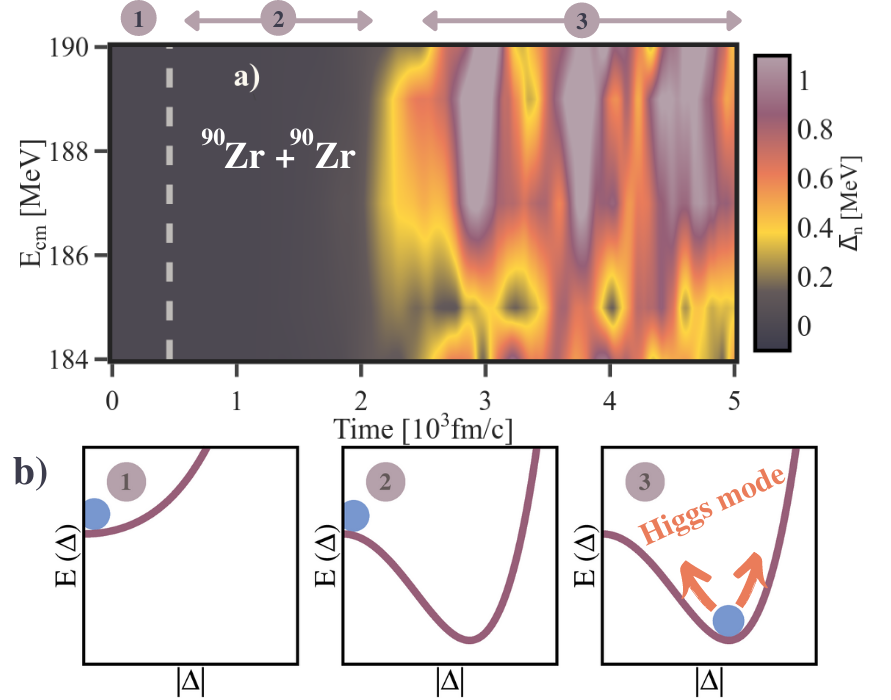}
\caption{Evolution of the average neutron pairing gap $\overline{\Delta}_{n}$ in collision $^{90}$Zr+$^{90}$Zr.
The moment of collision is denoted by the vertical dashed line. 
The upper panel corresponds to the case of $\Delta\phi=0$. 
The schematic figures in the lower panel describe the process of pairing Higgs mode excitation during collision. 
Each figure corresponds to a certain stage of the collision process which is 
indicated in the upper panel.
}
\label{fig:aVDnmap}
\end{figure}

In order to explore this effect the collision of $^{90}$Zr+$^{90}$Zr was considered for energies within a 10 MeV window at the Coulomb barrier. 
In order to be able to observe the symmetry breaking effect, the initial pairing field has to be kept as non-zero. 
Its average value was set to be small, $\overline{\Delta}_{n}=0.002$\,MeV.
The results are shown in Fig.~\ref{fig:aVDnmap}, where substantial pairing correlations, indicating symmetry breaking, set in at a certain time after collision (see also movies in the Supplemental Material\cite{SM}).
The pairing field exhibits relatively regular oscillation as expected in the case of the Higgs mode~\cite{behrle,bulgac19,bjerlin16}. 
The mechanism that leads to activation of this mode is schematically presented in Fig.~\ref{fig:aVDnmap}(b).
Before the collision the minimum of energy $E(|\Delta|)$ corresponds to neutron pairing gap $|\Delta|\approx 0$, since the colliding nuclei have a neutron magic number $N=50$. 
The composite system possesses an elongated structure and turns out to be dynamically unstable with respect to paring field fluctuations, which eventually triggers the Higgs oscillations.
Note that since the total energy is conserved, the energy gained by developing pairing correlations has to be transferred to other degrees of freedom.
For large amplitude oscillations the Higgs mode frequency is in range $|\Delta|\lesssim\hbar\omega_\text{H}\lesssim 2|\Delta|$~\cite{bulgacHiggs}.
For $|\Delta|\approx 1$\,MeV it gives the oscillation period in the range 600-1200\,fm/$c$, which agrees well with the observed oscillation period (especially well seen for $\Ecm\approx188$\,MeV).   

The induced pairing-field oscillation mode requires further investigation, as typically the Higgs mode is considered in the context of uniform systems, while here we have a finite and nonuniform case.  
Still, the results presented in this paper raise a justified doubt whether the description of collision of magic nuclei at low energy can be reliably described within frameworks that neglect pairing, e.g., the TDHF approach.
Another question which naturally arises is, what are observable consequences of the effect?
If the timescale for symmetry breaking is short enough, the effect should be pronounced at energies below the threshold in collisions of magic nuclei, as it should considerably affect TKE and TXE. 

\ssec{Conclusions}
We have shown that the effect of increasing the barrier in collisions of two superfluid nuclei with different phases of the pairing field is still present when realistic functionals with spin-orbit terms are used. 
The effective barrier increases with the magnitude of the pairing field.
We have confirmed that the effect is of purely dynamical origin related to spatially localized modulation of the pairing field, which we call a solitonic excitation.
The effective nucleus-nucleus repulsive interaction induced by the soliton suppresses significantly the excitation energy of the system.
The dynamics of the pairing field alter the shape evolution of the colliding system and enhance the role of the neck in the collision process. 
This effect may lead to modification of the dynamics of peripheral collisions where the interplay between the centrifugal force and the neck formation may affect the evolution.
We have shown that dynamics of the collision trigger the symmetry breaking and thus an enhancement of the magnitude of the pairing field. 
The mechanism of this effect can be traced back to modification of the density of states in the composite system leading to effective increase of the pairing effects.

\begin{acknowledgments}
We would like to thank Shi Jin and Ionel Stetcu for help concerning numerical issues related to TDSLDA.
This work was supported by the Polish National Science Center (NCN) under Contracts No. UMO-2016/23/B/ST2/01789 (P.M. and A.M.) and No. UMO-2017/26/E/ST3/00428 (GW).
We also acknowledge the Center for Computational Sciences (CCS), University of Tsukuba for resources at Cygnus (Project Name: TDSLDA20), the Global Scientific Information and Computing Center (GSICC), Tokyo Institute of Technology for resources at TSUBAME3.0 (Project No: hp210079), and the Interdisciplinary Centre for Mathematical and Computational Modelling (ICM) of Warsaw University for computing resources at Okeanos (Grant No. GA83-9). 
We also acknowledge PRACE for awarding us access to resource Piz Daint based in Switzerland at the Swiss National Supercomputing Centre (CSCS), Decision No. 2019215113.
\end{acknowledgments}


\sepline{}
\section*{Supplemental Material for: Pairing dynamics and solitonic excitations in collisions of medium-mass, identical nuclei}

Each movie consists of two panels. 
In the upper panel the density distributions are shown, while in the lower panels the absolute values of pairing field is presented.
Both density distributions and pairing fields are shown as sections along the symmetry axis.
In each panel, the upper (lower) part shows results for protons (neutrons).
Below the list movies is provided for various collision energies.
The value of the collision energy is encoded in the file name.

\subsection{$^{90}$Zr+$^{90}$Zr}
\subsubsection{$\bar{\Delta}_n \approx 0.01MeV, \Delta\phi=0$}
\begin{enumerate}
\itemovie{90Zr+90Zr\_SkM\_gn260\_0-Phase\_175MeV.mp4}{https://youtu.be/mySvL60b0dQ}
\itemovie{90Zr+90Zr\_SkM\_gn260\_0-Phase\_180MeV.mp4}{https://youtu.be/A6YxHYhE-xw}
\itemovie{90Zr+90Zr\_SkM\_gn260\_0-Phase\_182MeV.mp4}{https://youtu.be/mgmqq-q-IxA}
\itemovie{90Zr+90Zr\_SkM\_gn260\_0-Phase\_183MeV.mp4}{https://youtu.be/umDEcTDt8EY}
\itemovie{90Zr+90Zr\_SkM\_gn260\_0-Phase\_184MeV.mp4}{https://youtu.be/KW0Ut2grwl0}
\itemovie{90Zr+90Zr\_SkM\_gn260\_0-Phase\_185MeV.mp4}{https://youtu.be/Evg8swQ4fZo}
\itemovie{90Zr+90Zr\_SkM\_gn260\_0-Phase\_187MeV.mp4}{https://youtu.be/iq9QTvGU8pE}
\itemovie{90Zr+90Zr\_SkM\_gn260\_0-Phase\_189MeV.mp4}{https://youtu.be/EQQWuUasfzw}
\itemovie{90Zr+90Zr\_SkM\_gn260\_0-Phase\_190MeV.mp4}{https://youtu.be/lKpNCLw7LXs}
\itemovie{90Zr+90Zr\_SkM\_gn260\_0-Phase\_193MeV.mp4}{https://youtu.be/RbVw4aJRG8k}
\itemovie{90Zr+90Zr\_SkM\_gn260\_0-Phase\_196MeV.mp4}{https://youtu.be/cc-ms7yTFXw}
\itemovie{90Zr+90Zr\_SkM\_gn260\_0-Phase\_200MeV.mp4}{https://youtu.be/9PUeyFUQMXY}
\end{enumerate}

\subsubsection{$\bar{\Delta}_n \approx 0.01MeV, \Delta\phi=\pi$}
\begin{enumerate}
\itemovie{90Zr+90Zr\_SkM\_gn260\_PI-Phase\_175MeV.mp4}{https://youtu.be/x\_uecVvzU\_0}
\itemovie{90Zr+90Zr\_SkM\_gn260\_PI-Phase\_180MeV.mp4}{https://youtu.be/oIOLHMZU3h0}
\itemovie{90Zr+90Zr\_SkM\_gn260\_PI-Phase\_182MeV.mp4}{https://youtu.be/ib8BKTmKRbA}
\itemovie{90Zr+90Zr\_SkM\_gn260\_PI-Phase\_183MeV.mp4}{https://youtu.be/LA3W6tUuAVg}
\itemovie{90Zr+90Zr\_SkM\_gn260\_PI-Phase\_184MeV.mp4}{https://youtu.be/S1siBeVcfcs}
\itemovie{90Zr+90Zr\_SkM\_gn260\_PI-Phase\_185MeV.mp4}{https://youtu.be/4bFXVtX2-L0}
\itemovie{90Zr+90Zr\_SkM\_gn260\_PI-Phase\_187MeV.mp4}{https://youtu.be/2h7wws4djWs}
\itemovie{90Zr+90Zr\_SkM\_gn260\_PI-Phase\_189MeV.mp4}{https://youtu.be/25quUtxdjCE}
\itemovie{90Zr+90Zr\_SkM\_gn260\_PI-Phase\_190MeV.mp4}{https://youtu.be/8-iP-u8gK0g}
\itemovie{90Zr+90Zr\_SkM\_gn260\_PI-Phase\_193MeV.mp4}{https://youtu.be/wrrSk64iu8M}
\itemovie{90Zr+90Zr\_SkM\_gn260\_PI-Phase\_196MeV.mp4}{https://youtu.be/hyZNdfS0Lvw}
\itemovie{90Zr+90Zr\_SkM\_gn260\_PI-Phase\_200MeV.mp4}{https://youtu.be/\_Etm6IexKd0}
\end{enumerate}

\subsection{$^{96}$Zr+$^{96}$Zr}

\subsubsection{$\bar{\Delta}_n = 1.98MeV, \Delta\phi=0$}
\begin{enumerate}
\itemovie{96Zr+96Zr\_SkM\_gn280\_0-Phase\_175MeV.mp4}{https://youtu.be/ygKy2W27NJU}
\itemovie{96Zr+96Zr\_SkM\_gn280\_0-Phase\_178MeV.mp4}{https://youtu.be/peRCv0SP9L8}
\itemovie{96Zr+96Zr\_SkM\_gn280\_0-Phase\_179MeV.mp4}{https://youtu.be/9U4o0uMGZkE}
\itemovie{96Zr+96Zr\_SkM\_gn280\_0-Phase\_180MeV.mp4}{https://youtu.be/VUOAnFsIO10}
\itemovie{96Zr+96Zr\_SkM\_gn280\_0-Phase\_183MeV.mp4}{https://youtu.be/hqkE0u0-rkA}
\itemovie{96Zr+96Zr\_SkM\_gn280\_0-Phase\_185MeV.mp4}{https://youtu.be/NRnwiKeeqX0}
\itemovie{96Zr+96Zr\_SkM\_gn280\_0-Phase\_187MeV.mp4}{https://youtu.be/Jb4fatcyqmg}
\itemovie{96Zr+96Zr\_SkM\_gn280\_0-Phase\_190MeV.mp4}{https://youtu.be/JLLfrktqmsE}
\itemovie{96Zr+96Zr\_SkM\_gn280\_0-Phase\_193MeV.mp4}{https://youtu.be/Wwm06H6IIMw}
\itemovie{96Zr+96Zr\_SkM\_gn280\_0-Phase\_196MeV.mp4}{https://youtu.be/1Szp3Z3Jpb0}
\itemovie{96Zr+96Zr\_SkM\_gn280\_0-Phase\_200MeV.mp4}{https://youtu.be/JGH8Uo\_gXUo}
\end{enumerate}
\subsubsection{$\bar{\Delta}_n = 1.98MeV, \Delta\phi=\pi$}
\begin{enumerate}
\itemovie{96Zr+96Zr\_SkM\_gn280\_PI-Phase\_175MeV.mp4}{https://youtu.be/FgEF6A0a42k}
\itemovie{96Zr+96Zr\_SkM\_gn280\_PI-Phase\_178MeV.mp4}{https://youtu.be/nfCbIhjFKq4}
\itemovie{96Zr+96Zr\_SkM\_gn280\_PI-Phase\_179MeV.mp4}{https://youtu.be/KK9IWnBhpkE}
\itemovie{96Zr+96Zr\_SkM\_gn280\_PI-Phase\_180MeV.mp4}{https://youtu.be/BW5kF0m87Os}
\itemovie{96Zr+96Zr\_SkM\_gn280\_PI-Phase\_182MeV.mp4}{https://youtu.be/2xAVS3ucY1k}
\itemovie{96Zr+96Zr\_SkM\_gn280\_PI-Phase\_183MeV.mp4}{https://youtu.be/JILM-qjNahc}
\itemovie{96Zr+96Zr\_SkM\_gn280\_PI-Phase\_184MeV.mp4}{https://youtu.be/U7qupC-J\_Oc}
\itemovie{96Zr+96Zr\_SkM\_gn280\_PI-Phase\_185MeV.mp4}{https://youtu.be/Qvc5K3cZaOY}
\itemovie{96Zr+96Zr\_SkM\_gn280\_PI-Phase\_187MeV.mp4}{https://youtu.be/pKGi4dyrdgA}
\itemovie{96Zr+96Zr\_SkM\_gn280\_PI-Phase\_190MeV.mp4}{https://youtu.be/OCRtjkSyVns}
\itemovie{96Zr+96Zr\_SkM\_gn280\_PI-Phase\_193MeV.mp4}{https://youtu.be/KRaRzly2-ZE}
\itemovie{96Zr+96Zr\_SkM\_gn280\_PI-Phase\_196MeV.mp4}{https://youtu.be/ZWGDPnrWGZ8}
\itemovie{96Zr+96Zr\_SkM\_gn280\_PI-Phase\_200MeV.mp4}{https://youtu.be/hvU9uIe16KE}
\end{enumerate}

\subsubsection{$\bar{\Delta}_n = 2.44MeV, \Delta\phi=0$}
\begin{enumerate}
\itemovie{96Zr+96Zr\_SkM\_gn300\_0-Phase\_177MeV.mp4}{https://youtu.be/az-pRW08gEI}
\itemovie{96Zr+96Zr\_SkM\_gn300\_0-Phase\_178MeV.mp4}{https://youtu.be/k5bNybqvRRc}
\end{enumerate}
\subsubsection{$\bar{\Delta}_n = 2.44MeV, \Delta\phi=\pi$}
\begin{enumerate}
\itemovie{96Zr+96Zr\_SkM\_gn300\_PI-Phase\_177MeV.mp4}{https://youtu.be/36Wz9ejC2aw}
\itemovie{96Zr+96Zr\_SkM\_gn300\_PI-Phase\_178MeV.mp4}{https://youtu.be/1d7GNVMYb-c}
\itemovie{96Zr+96Zr\_SkM\_gn300\_PI-Phase\_184MeV.mp4}{https://youtu.be/eRspophIX\_I}
\itemovie{96Zr+96Zr\_SkM\_gn300\_PI-Phase\_185MeV.mp4}{https://youtu.be/Ue5cPoXJU4Q}
\itemovie{96Zr+96Zr\_SkM\_gn300\_PI-Phase\_186MeV.mp4}{https://youtu.be/EA0PDqMEJCo}
\itemovie{96Zr+96Zr\_SkM\_gn300\_PI-Phase\_187MeV.mp4}{https://youtu.be/0ilqHHxS1eM}
\end{enumerate}

\subsubsection{$\bar{\Delta}_n = 2.94MeV, \Delta\phi=0$}
\begin{enumerate}
\itemovie{96Zr+96Zr\_SkM\_gn320\_0-Phase\_170MeV.mp4}{https://youtu.be/bpCZgTtEzHU}
\itemovie{96Zr+96Zr\_SkM\_gn320\_0-Phase\_175MeV.mp4}{https://youtu.be/nQgstanTWSw}
\itemovie{96Zr+96Zr\_SkM\_gn320\_0-Phase\_177MeV.mp4}{https://youtu.be/bSr3O4ynTKs}
\itemovie{96Zr+96Zr\_SkM\_gn320\_0-Phase\_178MeV.mp4}{https://youtu.be/sI20famR\_VI}
\itemovie{96Zr+96Zr\_SkM\_gn320\_0-Phase\_180MeV.mp4}{https://youtu.be/r9uAsp8qi9g}
\itemovie{96Zr+96Zr\_SkM\_gn320\_0-Phase\_183MeV.mp4}{https://youtu.be/g2R\_idx5eOs}
\itemovie{96Zr+96Zr\_SkM\_gn320\_0-Phase\_185MeV.mp4}{https://youtu.be/UcDeEzsc0xk}
\itemovie{96Zr+96Zr\_SkM\_gn320\_0-Phase\_187MeV.mp4}{https://youtu.be/KR\_2fMKcHM8}
\itemovie{96Zr+96Zr\_SkM\_gn320\_0-Phase\_190MeV.mp4}{https://youtu.be/jSwUGZD0u6k}
\itemovie{96Zr+96Zr\_SkM\_gn320\_0-Phase\_193MeV.mp4}{https://youtu.be/2kwDm4JVeEE}
\itemovie{96Zr+96Zr\_SkM\_gn320\_0-Phase\_197MeV.mp4}{https://youtu.be/XKI7IG\_OUOU}
\itemovie{96Zr+96Zr\_SkM\_gn320\_0-Phase\_200MeV.mp4}{https://youtu.be/LAqb9ybmTY0}
\end{enumerate}
\subsubsection{$\bar{\Delta}_n = 2.94MeV, \Delta\phi=\pi$}
\begin{enumerate}
\itemovie{96Zr+96Zr\_SkM\_gn320\_PI-Phase\_175MeV.mp4}{https://youtu.be/ZSnaCe5obgM}
\itemovie{96Zr+96Zr\_SkM\_gn320\_PI-Phase\_177MeV.mp4}{https://youtu.be/VJ\_GSKHRZYM}
\itemovie{96Zr+96Zr\_SkM\_gn320\_PI-Phase\_178MeV.mp4}{https://youtu.be/s0ftdTjQQ0w}
\itemovie{96Zr+96Zr\_SkM\_gn320\_PI-Phase\_184MeV.mp4}{https://youtu.be/9\_8Kx2IbIuA}
\itemovie{96Zr+96Zr\_SkM\_gn320\_PI-Phase\_185MeV.mp4}{https://youtu.be/DZ5iUovIoS0}
\itemovie{96Zr+96Zr\_SkM\_gn320\_PI-Phase\_186MeV.mp4}{https://youtu.be/xCsD--SUyBo}
\itemovie{96Zr+96Zr\_SkM\_gn320\_PI-Phase\_187MeV.mp4}{https://youtu.be/ewN0KWyF6FU}
\itemovie{96Zr+96Zr\_SkM\_gn320\_PI-Phase\_188MeV.mp4}{https://youtu.be/46LGilXnGdg}
\itemovie{96Zr+96Zr\_SkM\_gn320\_PI-Phase\_190MeV.mp4}{https://youtu.be/g0ATBoce460}
\itemovie{96Zr+96Zr\_SkM\_gn320\_PI-Phase\_193MeV.mp4}{https://youtu.be/R4le7CkNzDQ}
\itemovie{96Zr+96Zr\_SkM\_gn320\_PI-Phase\_197MeV.mp4}{https://youtu.be/EqfTXBHe8F0}
\itemovie{96Zr+96Zr\_SkM\_gn320\_PI-Phase\_200MeV.mp4}{https://youtu.be/u-xM8lfPZU8}
\end{enumerate}

\end{document}